\documentclass[referee]{raa}            
\usepackage{graphicx,times}             
\input{epsf.sty}                        
\input{psfig.sty}                       

\begin{document}

\title{Multicolour Photometric Study of M31 Globular Clusters
}

   \volnopage{Vol.0 (2009) No.0, 000--000}      
   \setcounter{page}{1}          

   \author{Z. Fan
      \inst{1,2}
   \and J. Ma
      \inst{1}
   \and X. Zhou
      \inst{1}
   }

   \institute{National Astronomical Observatories, Chinese Academy of Sciences,
     20A Datun Road, Chaoyang District, Beijing 100012, China;
     {\it fanzhou@bac.pku.edu.cn}\\
     \and
     Graduate University of Chinese Academy of Sciences,
     19A Yuquan Road, Shijingshan District, Beijing 100049, China\\
   }

   \date{Received~~2009 month day; accepted~~2009~~month day}

\abstract{We present the photometry of 30 globular clusters (GCs)
and GC candidates in 15 intermediate-band filters covering from
$\sim$3000 to $\sim$10000 \AA~~using the archival CCD images of M31
observed as part of the Beijing - Arizona - Taiwan - Connecticut
(BATC) Multicolour Sky Survey. We transform these intermediate-band
photometric data to the photometry in the standard $UBVRI$
broad-bands. These M31 GC candidates are selected from the Revised
Bologna Catalogue (RBC V.3.5), and most of these candidates do not
have any photometric data. Therefore the present photometric data
are supplement to RBC V.3.5. We find that 4 out of 61 GCs and GC
candidates in RBC V.3.5 do not show any signal on the BATC images at
their locations. By linear fit of the distribution in
colour-magnitude diagram of blue GCs and GC candidates using the
data from RBC V.3.5, and in this study we find the ``blue-tilt'' of
blue M31 GCs with a high confidence at 99.95\% or $3.47\sigma$ for
the confirmed GCs, and $>99.99\%$ or $4.87\sigma$ for GCs and GC
candidates. }

   \authorrunning{Z. Fan, J. Ma \& X. Zhou }            
   \titlerunning{New Photometry of M31 GCs}  

   \maketitle

%
%
\section{Introduction}           
\label{sec:int} Globular clusters (GCs) are the oldest bound stellar
systems in the galaxies, so they provide a fossil record of the
earliest stages of galaxy formation and evolution. In addition,
since the GCs are bright ($<M_V> < -7.5$), they can be detected at a
large distance such as Virgo (see, e.g., Peng et al. ~\cite{peng06})
and Coma Clusters (see, e.g., Baum et al.~\cite{baum95}).
Furthermore, Kalirai et al.~(\cite{kali08}) discovered the GC system
(GCS) of a galaxy which is $\sim$375 Mpc far away ($z=0.089$) and
Mieske et al.~(\cite{mie04}) detected a GC system in Abell 1689
($z=0.183$), both of which are based on the deep image observations
with the Advanced Camera for Surveys (ACS) on $Hubble$ $Space$
$Telescope$ ($HST$). We can study the evolutionary process of the
distant galaxy through the nature of its GCs. Finally, GCs are
helpful to study the simple stellar population, since the
populations in a GC are generally thought to have the same age and
the same metallicity. However, some Galactic GCs, such as NGC~ 2808
and NGC~1851 are now suspected to be composed of heterogeneous
populations, and recent data from $HST$ are hinting at a great
fraction of Galactic GCs being composite populations at least
chemically (see, e.g., Yi ~\cite{yi09} and references there).

Located at a distance of $\sim780$ kpc (see, e.g., Stanek \&
Garnavich ~\cite{sg98}; Macri et al.~\cite{mac01}; McConnachie et
all.~\cite{mc05}), M31 is the largest and nearest Sb-type spiral
galaxy in Local Group. According to the latest catalogue: The
Revised Bologna Catalogue of M31 GCs and GC candidates (RBC V.3.5)
(Galleti et al.~\cite{gall04}; Galleti et al. ~\cite{gall06};
Galleti et al.~\cite{gall07}), there are 509 confirmed GCs and 1058
GC candidates discovered in M31 by far and 421 former GC candidates
have been proved to be stars, asterisms, galaxies, HII regions or
extended clusters. These GCs and GC candidates in RBC were observed
and discovered by many authors in different observation systems,
i.e. CCD photometry, photoelectric photometry, and photographic
plates, and a few are visual photometry (see, e.g.,
Vetesnik~\cite{vet62}; Sargent et al.~\cite{sar77}; Battistini et
al.~\cite{batt80}; Crampton et al.~\cite{cra85}; Barmby et
al.~\cite{barmby00}). For obtaining homogeneous photometric data,
Galleti et al.~(\cite{gall04}) took the photometric data of Barmby
et al.~(\cite{barmby00}) as reference and transformed others to this
reference and make the Master Catalogue RBC (see details in Galleti
et al.~\cite{gall04}). Although this catalogue includes the most
comprehensive photometry by far, there are dozens of GCs and GC
candidates nearly having not any photometric data. So it is
impending to present the photometry for these GCs and candidates.

In the present study we will present the BATC multicolour
photometric data for 30 GCs and GC candidates in RBC V.3.5, which
nearly have not any photometric and spectroscopic information.

This paper is organized as follows. In Sect. \ref{sec:saobs} we
present the BATC observations and Sect. \ref{sec:red} describes the
data reduction process. In Sect. \ref{sec:res} the final photometric
results are given and we compare our photometry to photometry in
literature. The new magnitude and colour distributions of the M31
GCs and GC candidates are also shown in Sect. \ref{sec:res}. And
finally, concluding remarks are given in Sect. \ref{sec:con}.

\section{The sample and observations}
\label{sec:saobs}

\subsection{The Sample GC Candidates}
\label{sec:sam}

To date, the study of M31 GCs has been mainly based on the
comprehensive Bologna Catalogue (Battistini et al.~
\cite{batt80,batt87,batt93}). Especially, Galleti et al.~
(\cite{gall04}) collected and revised all the available photometry,
obtaining homogeneous photometric data in $UBVRI$ bands by comparing
all the data with the CCD photometries of Barmby et
al.~(\cite{barmby00}) as a reference. In addition, Galleti et
al.~(\cite{gall04}) have searched the counterparts of the objects in
their catalogue based on the 2MASS database, providing integrated
$J, H, K_{s}$ photometry for 529 GCs and GC candidates, of which no
previous NIR photometry was observed. This catalogue is referred as
the Revised Bologna Catalogue of M31 globulars (hereafter RBC)
(Galleti et al.~ \cite{gall04}). It is worth mentioning that RBC is
frequently revised (Galleti et
al.~\cite{gall04,gall05,gall06,gall07}). The latest RBC is updated
on March 27, 2008, and referred as RBC V.3.5 \footnote{\em Please
see the details and download the catalogue from
http://www.bo.astro.it/M31/}, which includes the newly discovered
star clusters from Mackey et al.~(\cite{mac06}); Kim et
al.~(\cite{kim07}) and Huxor et al.~(\cite{hux08}). When we check
the distribution of $V$ mag in RBC V.3.5, we find that there are 61
GCs and GC candidates not having $V$ magnitudes, which are drawn in
Fig.\ref{fig:obs} with open and solid circles, and the solid circles
represent the multicolour photometries of these GCs and GC
candidates obtained in this paper. In fact, there are few
photometric data in any filters for these objects. So, in order to
obtain the photometry for these objects, we searched the BATC survey
archive during 1995 February $-$ 2008 March, i.e., from the
beginning to the end of M31 observation view before 2008 March
(since the M31 field cannot be observed in Xinglong Station after
March every year), covering about 6 square degrees, as showed in
Fig. \ref{fig:obs}. There are 519 individual images extracted, of
which the observation images during 1995 September $-$ 1999 December
have been dealt by Jiang et al.~ (\cite{jiang03}) (see their
Table~1). Table~\ref{tab:obs} lists the log of observations from
2000 to 2008. There are 7 of them (EXT8, SH25, B306D, DAO11, BA22,
BH01 and BA10) out of the BATC observation fields. So, the final
sample of M31 GC candidates of this paper includes 54 objects.
However, in this paper, we obtained the BATC multicolour photometry
for 30 objects. The photometries for the other 24 objects are not
obtained in this paper because of low signal-to-noise ratio or other
reasons, which will be discussed in detail below. By comparing with
Tables 1, 3, 4 and 5 of Caldwell et al. ~(\cite{cald09}), who
present a new catalogue of 670 likely star clusters, stars, possible
stars and galaxies in the field of M31, all with updated
high-quality coordinates being accurate to $0.2''$ based on the
images from the Local Group Survey (Massey et al.~ \cite{massey06})
or Digitized Sky Survey (DSS), we find that, of these 54 objects
only four objects (V234, H13, B523 and SK131C) are not included in
Tables 1, 3, 4 and 5 of Caldwell et al.~(\cite{cald09}). In
addition, there are 5 objects, the coordinates of which are
different between Galleti et al.~ (\cite{gall04}) (RBC V.3.5) and
Caldwell et al.~(\cite{cald09}). We listed them in
Table~\ref{Tab:coord2} for comparison. Since Caldwell et al.~
(\cite{cald09}) corrected the coordinates of M31 clusters based on
the Hectospec fibers and FK5 system, we used the coordinates in
Caldwell et al.~(\cite{cald09}) to obtain the multicolour
photometries for these 5 clusters in this paper.

Below, we will discuss the 24 objects in details, the photometries
of which were not obtained in this paper.

1. There are 13 GCs and GC candidates (B523, DAO32, NB27, NB31,
NB35, NB43, NB59, NB62, NB84, NB85, DAO83, SK131C and V229), the
signal-to-noise ratios of which are too low in the BATC survey
images of this paper, therefore we do not obtain their photometries.

2. NB57, NB60, SH05 and SH08: In the BATC survey images of this
paper, anything was not found on their positions given in RBC V.3.5.
Maybe, they are too faint to be seen in the BATC survey images.
However, there is an object very near the position of NB60 presented
by Galleti et al.~(\cite{gall04}). The R.A. and DEC of this object
are 00:42:26.63 and +41:18:04.5 compared to the R.A. and DEC of
00:42:26.68 and +41:18:10.70 presented by RBC V.3.5 for NB60.

3. B287: This cluster is very close to an object in the BATC survey
images (because of the low resolution of BATC system), the distance
of between the centers of these two objects is about 3 pixels in the
BATC images (in fact, from the BATC images, these two objects joint
together), and we cannot obtain its photometry accurately in this
paper.

4. B287D: It is very close to an object in the BATC survey images,
the distance of between the centers of these two objects is about 3
pixels in the BATC images, and we cannot obtain its photometry
accurately in this paper, too. In addition, this object is
classified as a star by Caldwell et al.~(\cite{cald09}).

5. B259D: This object is classified as a star by Caldwell et al.~
(\cite{cald09}). In addition, it overlaps another object in the BATC
survey images. Therefore, we did not obtain its photometry in this
paper, either.

6. DAO88: This object is faint, the magnitude in $V$ filter is
$19.82\pm0.15$ from Caldwell et al.~ (\cite{cald09}). In addition,
it is close to a much brighter star in the Milk Way (the magnitude
of this star in BATC $g$ filter is $12.23\pm0.002$). This star
contaminates DAO88, although the distance of the centers of these
two objects is 28 pixels in the BATC images, and we also cannot
obtain its photometry accurately in this paper. In addition, the
spectrum of this object is emission-line dominated from Caldwell et
al.~ (\cite{cald09}).

7. H13 and V300: In the BATC survey images, these two objects
overlap an object, respectively. So, we did not obtain its
photometry in this paper, either.

8. V254: It looks very extended in the BATC images, and is probably
not a star cluster. In fact, Caldwell et al.~(\cite{cald09}) has
classified this object as an HII region. Thus, we do not present its
multicolour photometry in this paper.

\subsection{Observations}
\label{sect:obs}

The observations of M31 were carried out by the BATC Multicolour Sky
Survey System, which uses a 60/90 cm f/3 Schmidt telescope at
Xinglong Station of the National Astronomical Observatories, Chinese
Academy of Sciences (NAOC), where the typical seeing condition is
$\sim2\arcsec-3\arcsec$ (Ma et al.~\cite{ma06a}). This system
includes 15 intermediate-band filters, covering a range of
wavelength from 3000 to 10000 \AA~(e.g., Fan et al.~\cite{fan96};
Zhou et al.~\cite{zhou03}). Before February 2006, a Ford Aerospace
$\rm{2k} \times \rm{2k}$ thick CCD camera was applied, which has a
pixel size of 15 $\mu\rm{m}$ and a field of view of $58^{\prime}$
$\times $ $58^{\prime}$, resulting in a resolution of
$1.7''\rm{pixel}^{-1}$. Later till now, a new E2V
$\rm{4k}\times\rm{4k}$ thinned CCD with a pixel size of 12
$\mu\rm{m}$, has successfully been mounted on the focal plane of the
Schmidt telescope, resulting in a resolution of
$1.3''\rm{pixel}^{-1}$. The blue quantum efficiency of the new E2V
thinned CCD, which is $92.2\%$ at 4000~\AA, is much higher than the
old thick one. Please refer Table~\ref{tab:ccd} for comparing the
parameters between thick and thinned CCD.

M31 fields were observed in several observing runs: the first run is
on the nights of UT 1995 February $-$ 1999 December; the second
observation is on the nights of UT 2004 January 12$-$27 and 2004
February 1$-$7; the third observation is on the nights of UT 2004
August 12$-$19, September 6$-$27 and October 1$-$12; the fourth
observation is on the nights of UT 2005 October 31 and November
1$-$7; the fifth observation is on the nights of UT 2006 October
6$-$29 and November 1$-$18. Table~\ref{tab:obs} summarizes the
observational catalogue including the name of observed field, filter
name, central wavelength, width of filter, numbers of image
combined, exposure time, and limiting magnitude, respectively.

\begin{figure}
\centering
\includegraphics[scale=0.5,angle=-90]{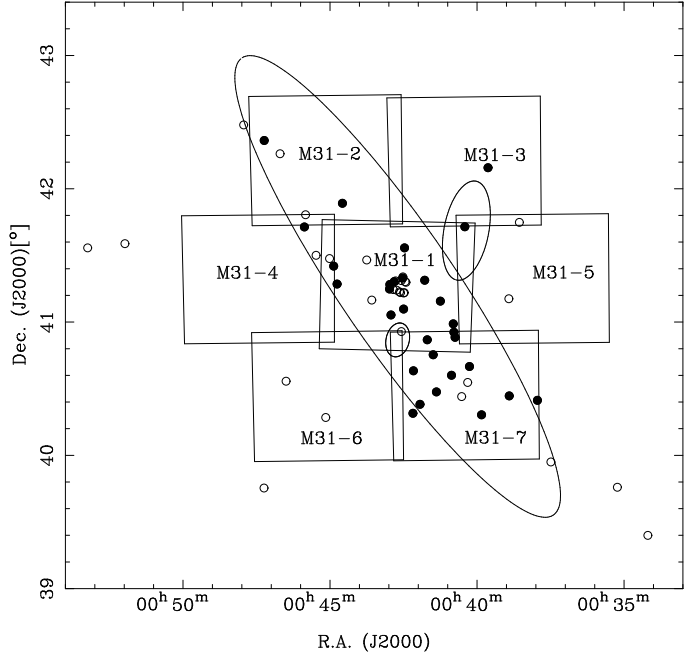}
\vspace{.3cm} \caption{The BATC observation of M31 field and every
field box is $58 \arcmin\times 58 \arcmin$ (for the old CCD). The
large ellipse is the boundary of disk and halo of M31
(Racine~\cite{rac91}). The other two small ellipses are $D_{25}$ of
NGC 205 (northwest) and M32 (southeast). All the symbols (both the
open and the solid) indicate 61 GCs and GC candidates not having $V$
magnitudes in RBC V.3.5, and the solid symbols represent that the
multicolour photometries of these GCs and GC candidates are obtained
in this paper.} \label{fig:obs}
\end{figure}

Fig. \ref{fig:obs} shows the location of 7 observed fields those are
drawn with boxes, the names of which are labeled in each box. The
total observed field of view is about 6 square degrees.

\section{Data Reduction and Photometry}
\label{sec:red}

Descriptions of the BATC photometric system can be found in Fan et
al.~(\cite{fan96}). Bias subtraction and flat-fielding with dome
flats were done with the BATC automatic data reduction software,
PIPELINE I, developed for the BATC Multicolour Sky Survey (Fan et
al.~\cite{fan96}; Zheng et al.~\cite{zheng99}). The dome flat-field
images were taken by using a diffuser plate in front of the
correcting plate of the Schmidt telescope, and the flatfielding
technique has been verified (see e.g., Fan et al.~\cite{fan96};
Zheng et al.~\cite{zheng99}; Wu et al.~\cite{wu02}; Yan et al.~
\cite{yan00}; Zhou et al.~\cite{zhou01,zhou04}). Spectrophotometric
calibration of the M31 images are made by observations of four $F$
sub-dwarfs, HD~19445, HD~84937, BD~${+26^{\circ}2606}$ and
BD~${+17^{\circ}4708}$, all taken from Oke \& Gunn~(\cite{og83}).
Hence, our magnitudes are defined in a way similar to the
spectrophotometric AB magnitude system (i.e, the Oke \& Gunn
$\tilde{f_{\nu}}$ monochromatic system).

The BATC magnitudes (see, e.g., Yan et al.~\cite{yan00}; Zhou et
al.~ \cite{zhou01,zhou03}) of the AB magnitude system is defined as

\begin{equation}
m_{\rm BATC}=-2.5{log\frac{\int_{\lambda_{1}}^{\lambda_{2}}{d} ({\rm
log}\nu)f_{\nu}r_{\nu}} {\int_{\lambda_{1}}^{\lambda_{2}}{d} ({\rm
log}\nu)r_{\nu}}}-48.60,
\end{equation}

\noindent which links the magnitude to the number of photons
detected by the CCD rather than to the input flux (Fukugita et al.~
\cite{fuku96}). In equation (1), $\nu$ is frequency; $f_{\nu}$ is
the spectral energy distribution of the source in unit of $\rm
erg\,s^{-1}\,cm^{-2}\,Hz^{-1}$; $r_{\nu}$ is the filter response
function of the system; $\lambda_{1}$ and $\lambda_{2}$ are the
lower and upper cutoff wavelength of the passband, respectively.

\subsection{Calibrations for $a$ and $b$ Intermediate-band Filters
of M31-1 Field}
\label{sec:T523ab}

Jiang et al.~(\cite{jiang03}) and Ma et al.~(\cite{ma06b, ma06c,
ma07}) studied 203 M31 GCs and GC candidates based on the BATC
observations in $c$ to $p$ intermediate-band filters for the M31
central field (M31-1 in Fig. 1). In this paper, we also used the
combined images and calibrated results of Jiang et
al.~(\cite{jiang03}) to obtain the photometric data for our sample
GCs and GC candidates in the intermediate-band filters of $c$ to
$p$, which are in the M31-1 field. For the new observed images of
the M31-1 field in $a$ and $b$ intermediate-band filters, we reduced
them by automatic reduction software: Pipeline I, which includes
bias subtraction and flat fielding of the CCD images. Then, we
combined the images observed in the same filter to eliminate cosmic
rays and to increase the signal-to-noise ratios as usually done in
photometry. The absolute flux of the combined images was calibrated
using the observations of standard stars (see, e.g., Fan et
al.~\cite{fan96}; Zheng et al.~\cite{zheng99}; Wu et
al.~\cite{wu02}; Yan et al.~ \cite{yan00}; Zhou et
al.~\cite{zhou01,zhou04}).

Table~\ref{tab:T523ab} lists the observational parameters of the
BATC M31-1 filed in $a$ and $b$ filters: filter name, central
wavelength, width of filter, numbers of image combined, exposure
time, the calibration errors in magnitude of the standard stars, and
limiting magnitude, respectively.

\subsection{Calibrations of M31-2 to M31-7 Fields}
\label{sec:Tma16}

For the images of the M31-2 to M31-7 fields, we did the same imaging
reduction and combination as we did in Sect.\ref{sec:T523ab}. The
absolute flux of the combined images of the M31-2 to M31-7 fields
was calibrated based on secondary standard transformations using the
M31-1 field; we could easily identify the stars in common between
the M31-2 to M31-7 fields and the M31-1 field, since adjacent
overlapping fields were initially arranged. On the image of each
filter, we first identified the positions of the common stars in the
overlapping fields, calculated the mean magnitude offsets between
the standardized magnitudes and instrumental magnitudes, and then
applied this magnitude offset to transform the instrumental
magnitudes of the M31-2 to M31-7 fields to the standardized
magnitudes.

\subsection{IRAF/DAOPHOT Photometry}
\label{sec:pho}

\begin{figure}
\centering
\includegraphics[scale=0.6]{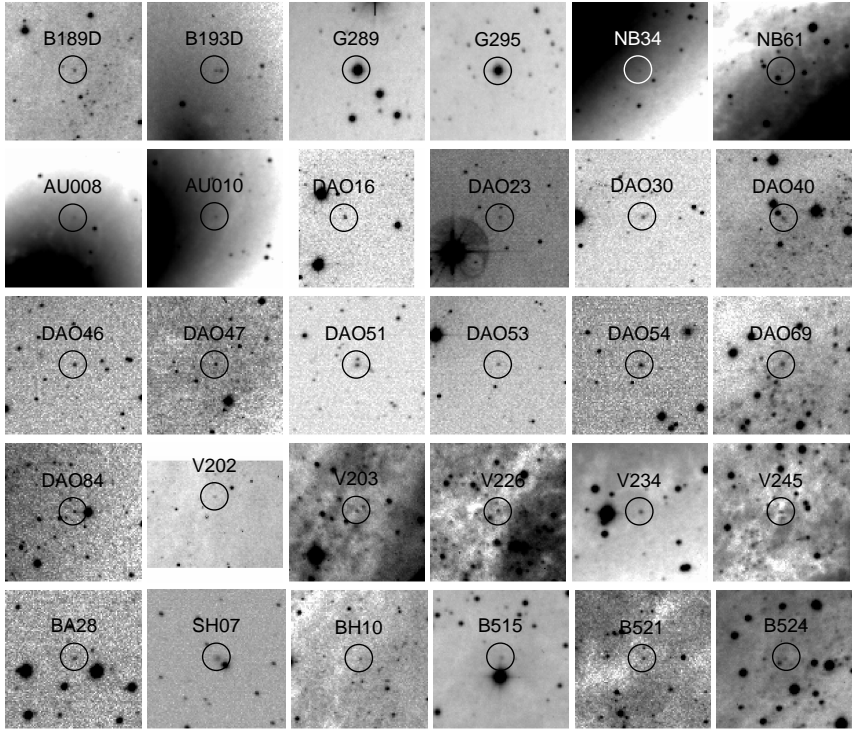}
\vspace{-7.0cm} \caption[]{Images of the sample GC candidates in the
BATC $g$ band obtained with the NAOC 60/90cm Schmidt telescope,
which are circled. The field of view of the image is $2.8 \times
2.8~ \rm {arcmin}^2$.} \label{fig:2}
\end{figure}

For each M31 GC and GC candidate, the PHOT routine in DAOPHOT
(Stetson~\cite{stet87}) is used to obtain magnitudes. To avoid
contamination from nearby objects, we adopt an aperture of a radius
of 3 pixels on the Ford CCD and of a radius of 4 pixels on the E2V
CCD. Inner and outer radius for background determination are taken
at 8 to 13 pixels for the 2k$\times$2k Ford CCD corresponding to 10
to 17 pixels for the 4k$\times$4k E2V CCD, from the center of the
objects. Given the small aperture use for the GC and GC candidate
observations, aperture corrections are determined as follows: We use
the isolated stars to determine the magnitude difference between
photometric radius of 3 pixels on the Ford CCD images and of 4
pixels on the E2V CCD images and the full magnitude of these stars
in each of the 15 BATC filters. The spectral energy distributions
(SEDs) for the GCs and GC candidates are then corrected for this
difference in each filter. However, for AU008, since its background
is too bright and varies greatly, we adopted an aperture radius of 2
pixels, and inner and outer radius for background determination are
taken at 3 to 6 pixels. The SEDs for the sample GCs and GC
candidates are listed in Table~\ref{tab:batc}. Column 2 to Column 16
give the magnitudes of the 15 BATC passbands observed. The second
line for each object gives the $1-\sigma$ errors in magnitudes for
the corresponding passband. The errors for each filter are given by
DAOPHOT. For some GCs and GC candidates, the magnitudes in some
filters could not be obtained due to low signal-to-noise ratio. Fig.
\ref{fig:2} shows the finding charts of the sample GCs and GC
candidates in this paper in the BATC $g$ band (centered on 5795\AA),
obtained with the NAOC 60/90cm Schmidt telescope.

\subsection{Transformed Magnitudes in Broadband System}

It is an uncontroverted fact that the photometry in the broadband
system is very commonly used in astrophysical studies. In RBC V.3.5,
the authors transformed the magnitudes of M31 605 GCs and GC
candidates from Kim et al.~(\cite{kim07}) in the Washington $CMT_1$
photometry to ones in broadband $UBVRI$ system based on the
equations of Geisler~(\cite{Geisler96}). These GCs and GC candidates
were searched on the CCD images observed by Kim et
al.~(\cite{kim07}) using the KPNO 0.9m telescope. In order to keep
consistency with RBC V.3.5, we will transform the magnitudes of the
sample GCs and GC candidates obtained in this paper to ones in
broadband $UBVRI$ system. Using Landolt standards and the catalogues
of Landolt~(\cite{land83,land92}) and of Galad\'\i-Enr\'\i quez,
Trullols \& Jordi~(\cite{gtj00}), Zhou et al.~(\cite{zhou03})
derived the relationships between the BATC intermediate-band system
and the $UBVRI$ broad-band system. These relationships are given in
equations (2) and (6) as:

\begin{equation}
U = m_b + 0.6801 (m_a - m_b) - 0.8982 \pm 0.143
\label{eq:U}
\end{equation}

\begin{equation}
B = m_d + 0.2201 (m_c - m_e) + 0.1278 \pm 0.076
\label{eq:B}
\end{equation}

\begin{equation}
V = m_g + 0.3292 (m_f - m_h) + 0.0476 \pm 0.027
\label{eq:V}
\end{equation}

\begin{equation}
R = m_i + 0.1036 \pm 0.055
\label{eq:R}
\end{equation}

\begin{equation}
I = m_o + 0.7190 (m_n - m_p) - 0.2994 \pm 0.064
\label{eq:I}
\end{equation}

$1-\sigma$ errors of the magnitudes were estimated using the
formulas below,

\begin{equation}
\sigma_{U} = \sqrt{{\sigma_b}^2 + 0.6801^2 (\sigma_a^2 + \sigma_b^2)}
\label{eq:Uerr}
\end{equation}

\begin{equation}
\sigma_{B} = \sqrt{{\sigma_d}^2 + 0.2201^2 (\sigma_c^2 + \sigma_e^2)}
\label{eq:Berr}
\end{equation}

\begin{equation}
\sigma_{V} = \sqrt{{\sigma_g}^2 + 0.3292^2 (\sigma_f^2 + \sigma_h^2)}
\label{eq:Verr}
\end{equation}

\begin{equation}
\sigma_{R} = \sigma_i
\label{eq:Rerr}
\end{equation}

\begin{equation}
\sigma_{I} = \sqrt{{\sigma_o}^2 + 0.7190^2 (\sigma_n^2 + \sigma_p^2)}
\label{eq:Ierr}
\end{equation}

Using these equations, we calculate the $UBVRI$ broadband magnitudes
and 1$-\sigma$ errors for the GCs and GC candidates in this paper
and list them in Table~\ref{tab:broad}.

\section{Results and Discussions}
\label{sec:res}

\subsection{Comparison of the Broadband Magnitude}

Very recently, Caldwell et al.~(\cite{cald09}) presented an updated
catalogue for 1300 objects in M31, including 670 likely star
clusters, and this catalogue presented the magnitudes in the $V$
band for the most objects based on the observations of LGS survey of
M31. In order to check the photometry in this paper, we compared the
results between Caldwell et al.~(\cite{cald09}) and this paper. In
Fig. 3, we plot the comparison of $V$ (BATC) photometry with the
measurement of Caldwell et al.~(\cite{cald09}). In this figure, our
magnitudes are on the x-axis, the difference between our and
Caldwell et al.~(\cite{cald09}) magnitudes are on the y-axis. The
mean $V$ magnitude difference (in the sense of this paper $-$
Caldwell et al.~(\cite{cald09})) is $<\Delta V> =-0.13\pm0.27$.

\begin{figure}
  \centering
  \includegraphics[scale=0.6,angle=-90]{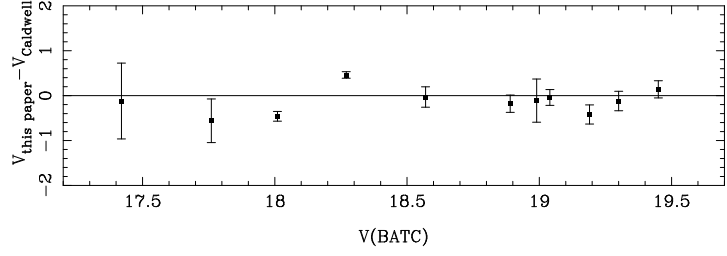}
  \caption{Comparison of GC candidate photometry with previous
    measurements by Caldwell et al.~(\cite{cald09}).}
  \label{fig:3}
\end{figure}

\subsection{Magnitude and Colour Distributions}
\label{sec:dis}

\begin{figure}
  \centering
  \includegraphics[scale=0.6]{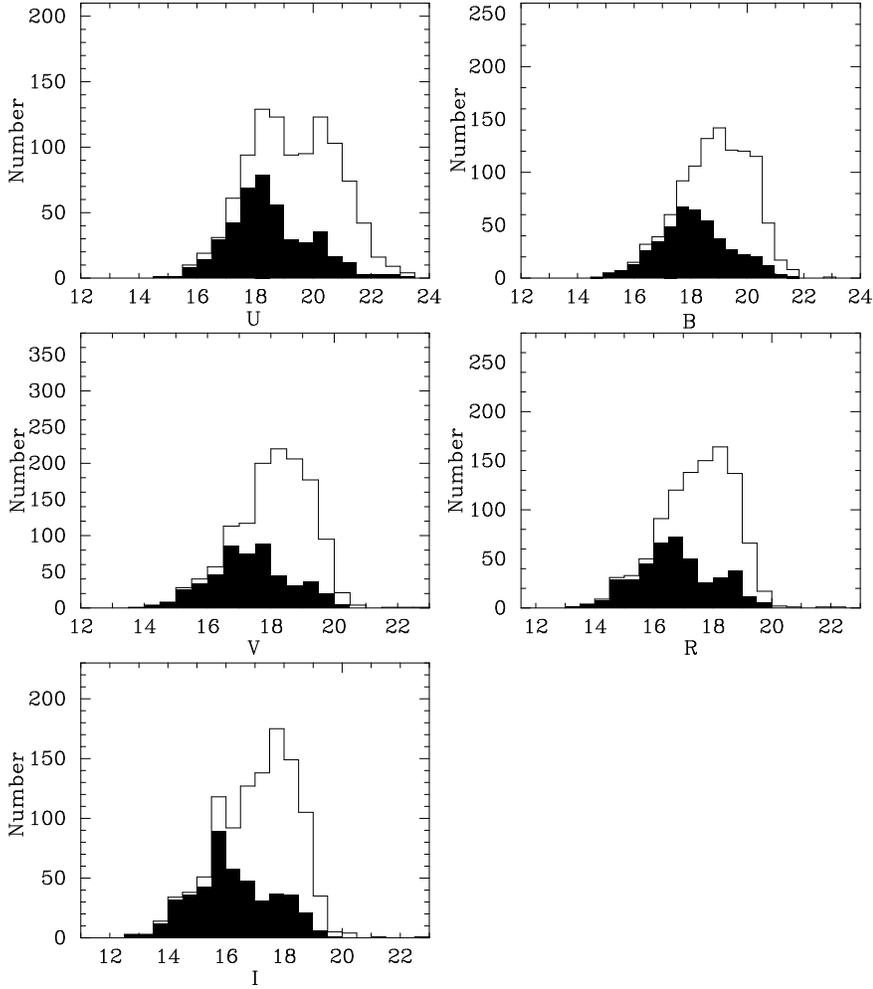}
  \caption{The distributions of the  magnitudes of M31 GCs and candidates
    in $U, B, V, R$ and $I$ bands, which combines the photometry from RBC V.3.5
    and those from Table~\ref{tab:broad}. The open histograms are for
    the GCs and GC candidates while the hatched histograms are for the
    confirmed GCs.}
  \label{fig:magdis}
\end{figure}

Based on the combined photometry of RBC V.3.5 and
Table~\ref{tab:broad} in this paper, Fig. \ref{fig:magdis} plots the
$UBVRI$ magnitude distributions of M31 confirmed GCs and GC
candidates classified in RBC V.3.5. The distributions for the
confirmed GCs ($f=1$ in RBC V.3.5) are presented with the hatched
histograms while the open histograms are for the GCs and GC
candidates ($f=1$, 2 and 3 in RBC V.3.5) \footnote{Note that there
are 265 GC candidates in RBC V.3.5, which are classified as stars
(148), galaxies (117) by Caldwell et al.~(\cite{cald09}). So, we did
not include these objects in Fig. 4}. We indicate that in RBC V.3.5,
$f=1$, 2 and 3 represent confirmed GCs, GC candidates and
controversial objects. The controversial objects, the total number
of which is only 9 in RBC V.3.5, mean that, for example, some
objects have been classified as galaxies based on high resolution
ground images, while they are classified as GCs on the basis of
spectroscopical observation (see, Galleti et al.~\cite{gall04} for
more details). They need to be confirmed in future. So they are
included as GC candidates in this paper. In all, combining the
photometry in RBC V.3.5 and this paper, there are totally 1029,
1085, 1289, 1033 and 1093 photometric values in the $U,B,V,R$ and
$I$ bands for GCs and GC candidates in M31, respectively, and there
are 428, 449, 506, 423 and 456 confirmed GCs having photometric data
in $U,B,V,R$ and $I$ bands, respectively. Obviously, the confirmed
GCs only occupy a small part of candidates. So, in the future, it is
also impending to confirm GCs from GC candidates.

In addition, Fig. 5 also plots the colour distributions for the GCs
and GC candidates with the same photometric data as in Fig.
\ref{fig:magdis}. For the $bona fide$ GCs (the hatched histograms),
the peak values of the colour distributions in $B-V$, $B-R$, $B-I$,
$V-R$, $U-B$ and $V-K$ are about 0.8, 1.3, 1.9, 0.5, 0.3 and 2.5,
respectively. The peak values of the colour distributions for all
the GCs and GC candidates are nearly the same as those of the $bona
fide$ GCs except for $V-K$ colour. It seems that the confirmed GCs
dominate the bluest part of all the GCs and GC candidates in the
$V-K$ colour distribution while the other colour distributions do
not show this phenomenon, which is also seen in Fig. \ref{fig:twoc}.
This may be due to the fact that Near Infrared (NIR) photometry of
GCs and GC candidates in RBC V.3.5 are actually lacked, which can be
easily seen based on the number of $V-K$ colours. 2MASS is the main
source of NIR photometry in RBC V.3.5. However, as the exposure time
is usually too short in 2MASS survey, the NIR photometries of faint
GCs and GC candidates are not obtained in RBC V.3.5. In addition,
bright GCs are easily searched.

\begin{figure}
  \centering
  \includegraphics[scale=0.6]{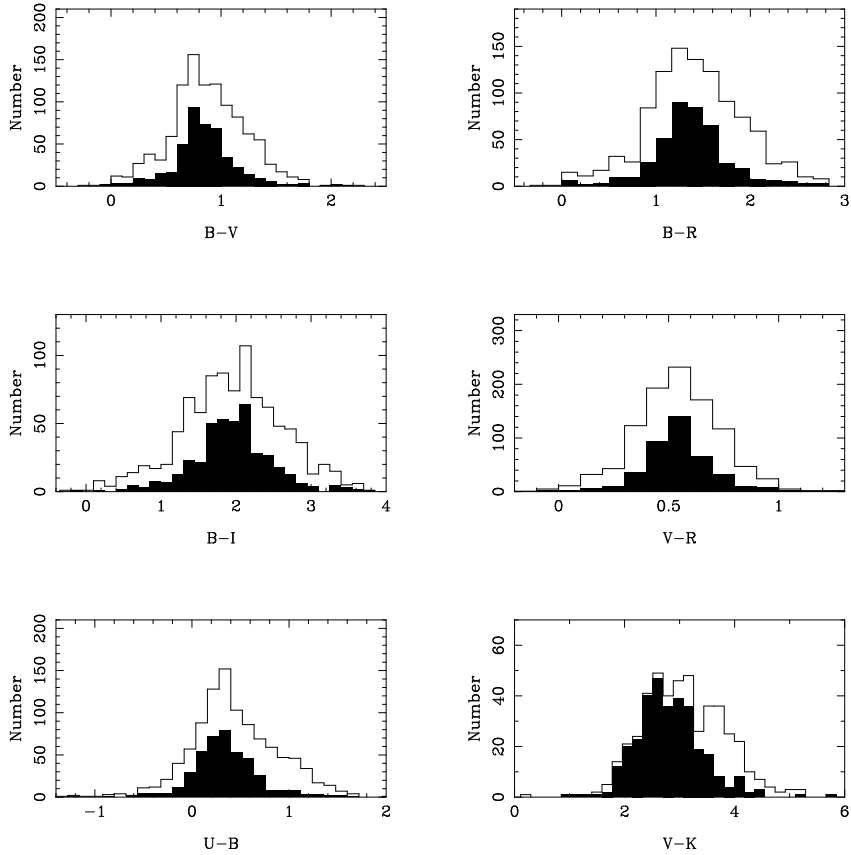}
  \caption{The colour distributions of M31 GCs and candidates, which
    combines the photometric data from RBC V.3.5 and this paper.
    The open solid histograms represent the GCs and candidates
    while the hatched histograms represent the confirmed GCs.}
  \label{fig:coldis}
\end{figure}

Fig. \ref{fig:twoc} plots the colour - colour diagrams of GCs and GC
candidates in RBC V.3.5: $V-K$ vs various colours, which are from
RBC V.3.5 and Table~\ref{tab:broad}. The confirmed GCs are marked
with red pluses while the GC candidates are shown as black pluses.
The reason that we use the $V-K$ in every colour-colour diagram, is
$V-K$ can provide a useful discriminant to separate the $bona fide$
GCs from the background galaxies (Galleti et al.~\cite{gall04}). In
this figure, it is easy to find that most of the $bona fide$ GCs
with $V-K <$ 3.0 while most of the $V-K >$ 3.0 points are GC
candidates. Therefore, there are a lot of GC candidates might be the
background galaxies according to the conclusion of Galleti et
al.~(\cite{gall04}).

\begin{figure}
  \centering
  \includegraphics[scale=0.6]{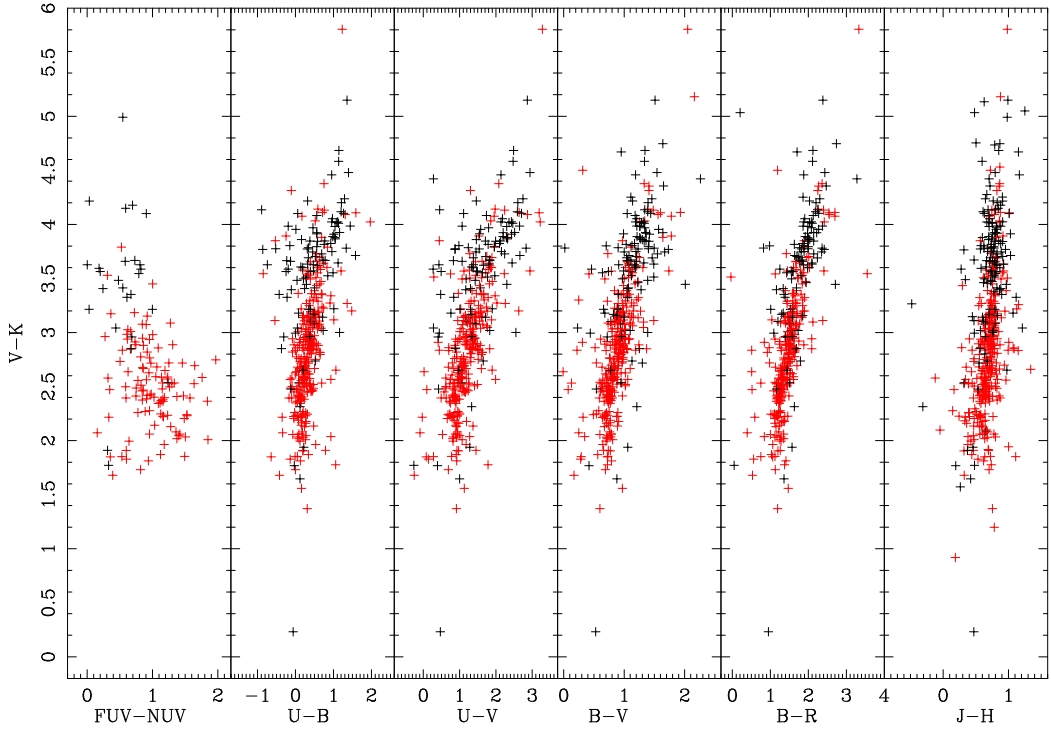}
  \caption{Colour-colour diagrams of GCs and GC candidates: $V-K$
    versus various colours. The photometric data are from RBC V.3.5 and
    this paper. The red pluses represent confirmed GCs while the black
    pluses represent GC candidates.}
  \label{fig:twoc}
\end{figure}

\begin{figure}
  \centering
  \includegraphics[scale=0.7]{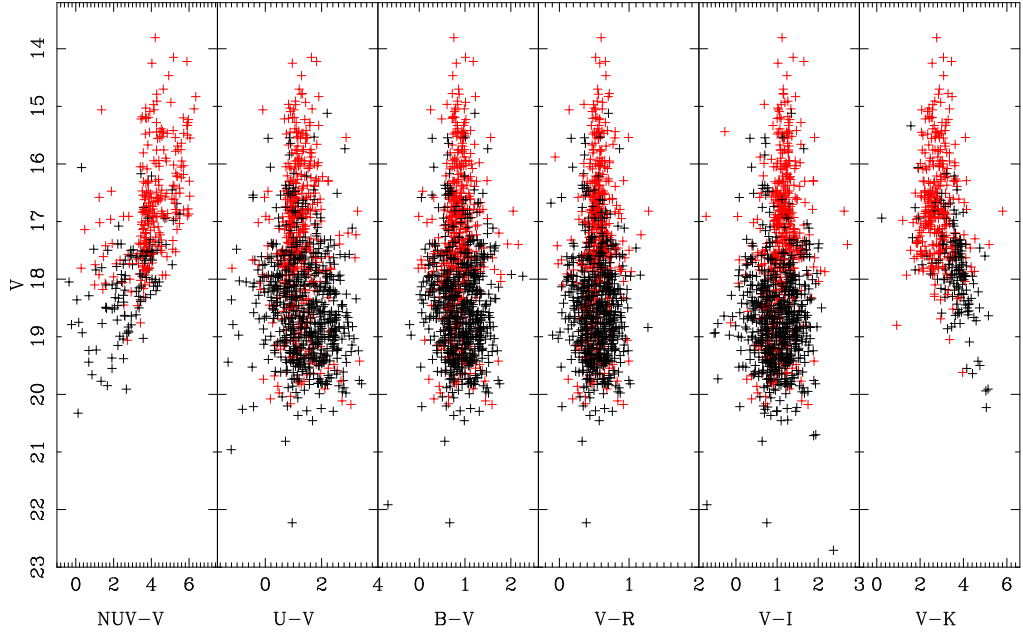}
  \caption{Colour - $V$ magnitude diagrams for the GCs and GC candidates.
    The photometric data are from RBC V.3.5 and Table~\ref{tab:broad}.
    The red pluses represent confirmed GCs while the black pluses
    represent GC candidates.}
  \label{fig:cmd}
\end{figure}

Fig. \ref{fig:cmd} shows the colour - $V$ magnitude diagrams for the
GCs and GC candidates. Most of the candidates are fainter than $V=$
17 while most of the bright objects ($V<$ 17) are confirmed GCs. It
implies that the fainter candidates are difficult to be identified and
the bright GCs are much easier to be searched.

\subsection{The Colour - Luminosity Relationship}

Harris et al.~(\cite{ha06}) investigated the colour magnitude
diagrams (CMDs) of GCs in eight brightest cluster galaxies (BCGs)
with the ACS/WFC data of {\sl HST}, and found a trend that the
redder GCs are more luminous (massive) for the blue (metal-poor)
population with $M_I > -10.5$, which is called ``blue tilt''. After
that, Strader et al.~ (\cite{str06}) also found the blue tilt
phenomenon in giant ellipticals (gEs) M87 and NGC 4649 in Virgo
Cluster by analyzing the {\sl HST}/ACS images, and used the
self-enrichment to interpret this phenomenon. Spitler et
al.~(\cite{spt06}) also utilized the {\sl HST}/ACS images to study
the Sa/S0 galaxy Sombrero (NGC 4594) and found the blue tilt in the
CMD. Mieske et al.~(\cite{mie06}) found that, the blue tilt exists
in early galaxies from the brightest ones to the faintest ones in
Virgo Cluster with the {\sl HST}/ACS observations, and the slope is
deeper in the more luminous host galaxies. Mieske et
al.~(\cite{mie06}) indicated that self-enrichment and field star
capture, or a combination of the two processes, offer the most
promising means of explaining the blue tilt. Strader \&
Smith~(\cite{ss08}) analyzed all the possible explanations for the
blue tilt phenomenon discovered previously in various external
galaxies, and showed that the self-enrichment in proto-GC clouds can
reproduce some aspects of the blue tilt. In this model, star
formation is controlled by supernova feedback, and the efficiency
scaling is proportional to the protocloud mass. Strader \&
Smith~(\cite{ss08}) also investigated the metallicity and mass
relationship of Galactic GCs, and did not find the blue tilt; and
they suggested that this might be due to the fundamental differences
between the parent clouds of Galactic GCs and those in blue-tilt
galaxies. However, Strader \& Smith~(\cite{ss08}) do not know
whether the blue tilt exits in M31 or not. We will investigate this
phenomenon based on the data in RBC V.3.5 and in this paper.

\begin{figure}
  \centering
  \includegraphics[scale=0.5]{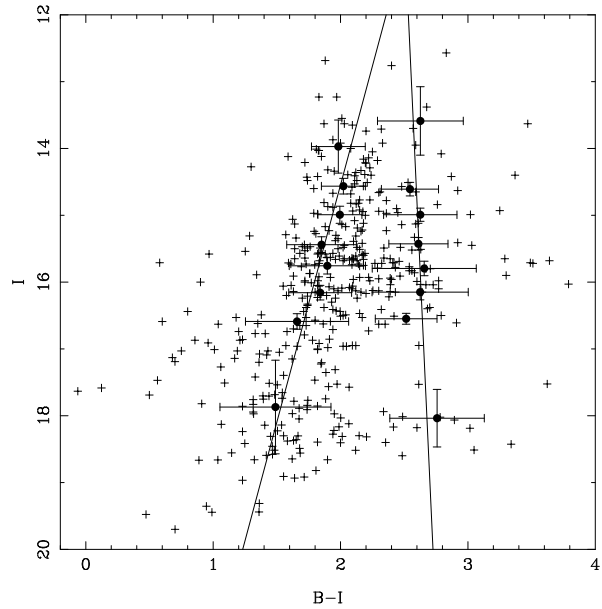}
  \caption{$I$ versus $B-I$ diagram for all the confirmed GCs in
    M31. The data used are from RBC V.3.5 and this paper.
    The linear fits show blue tilt for the blue population while
    there is no such relationship for the red population.}
    \label{fig:tilt1}
\end{figure}

\begin{figure}
  \centering
  \includegraphics[scale=0.5]{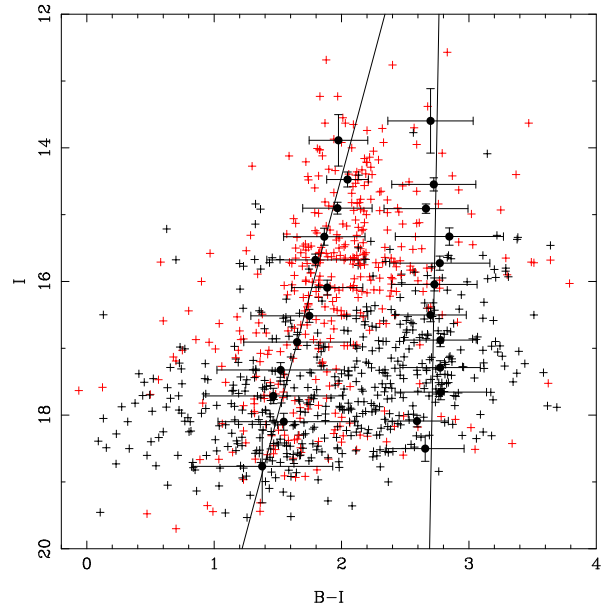}
  \caption{The same as Fig. \ref{fig:tilt1}, but for all the
  confirmed GCs (red pluses) and GC candidates (black pluses).
  The data used are also from RBC V.3.5 and this paper.
    We can also find the blue tilt in this figure.} \label{fig:tilt2}
\end{figure}

We plot the CMDs in ($B-I$) versus $I$ band for all the confirmed
GCs of M31 in Fig. \ref{fig:tilt1} and for all the confirmed GCs and
GC candidates of M31 in Fig. \ref{fig:tilt2}. The KMM mixture
modeling routine (Ashman, Bird \& Zepf~\cite{ash94}) was performed
to distinguish the blue and red subpopulations in both figures. Then
we divided the GCs into several mag bins
to calculate the mean colour and mean magnitude in each bin (see the
big black spots in Figs. \ref{fig:tilt1} and \ref{fig:tilt2}). We
fit the linear relationship for the colour $B-I$ and magnitude $I$
with the formula: $B-I=a+bI$. First we fit the data of the confirmed
GCs, for the blue population we have $b=-0.141\pm0.021$ while for
red population $b=0.024\pm0.020$; then for all the GCs and GC
candidates, we have $b=-0.140\pm0.013$ for the blue population and
$b=-0.009\pm0.014$ for the red population. The results strongly
present evidence of significant slopes for the blue population.
However, the slopes for the red population nearly approach zero, in
other words, the luminosity of the red GC population is completely
independent of its colour, which means there is no such relationship
for red GC population.

\section{Conclusions}
\label{sec:con}

In this paper, we obtain the  photometry of 30 M31 GCs and GC
candidates in 15 BATC intermediate bands. Based on Eqs. $6-10$ of
Zhou et al.~ (\cite{zhou03}), we transformed the BATC
intermediate-band magnitudes to the boradband $UBVRI$ magnitudes for
these GCs and GC candidates. We checked the objects in our BATC
images and found that, 4 objects could not be detected in the
locations presented in RBC V.3.5.

We also investigated the relationship between $I$ mag and $B-I$
colour for the confirmed GCs, and GCs and GC candidates,
respectively. After fitting, we obtained the slope of the blue
population $b=-0.141\pm0.021$ for the confirmed GCs, and
$b=-0.140\pm0.013$ for all the GCs and GC candidates.

Finally, we point out that our new supplementary photometry to RBC
V.3.5 is not only helpful for us to understand the total nature of
GCS in M31 but also useful for the Large Sky Area Multi-Object Fiber
Spectroscopic Telescope (LAMOST) project of China to estimate the
exposure time when observing them by LAMOST. LAMOST is being built
in Xionglong Station of National Astronomical Observatory, Chinese
Academy of Sciences (NAOC), Heibei province of P. R. China. This
telescope is a reflecting Schmidt telescope with an effective
aperture of 4 meters and a primary mirror of 6 meters mounted on it,
applying the active optics and fiber positioning technology. The
angle diameter of the field of view (FOV) is 5 degree and 4000
fibers are on the large focal plane of 1.75-meter in diameter, which
makes it possess the ability to produce several ten-thousands of
spectra in one night for the faint objects down to $V=20.5$ mag. The
LAMOST project has been completed in the end of 2008 and its
spectroscopic survey will make important contributions to stellar
astrophysics, the Galaxy, extra-galactic astrophysics and cosmology
(Zhao~\cite{zhao99}). Thus it is proper to use LAMOST to investigate
properties of GCs and GC candidates in M31 because of its wide FOV
(5 degree) and many star clusters can be observed at the same time.

LAMOST is appropriate to study the properties of M31 GCs and GC
candidates: in a few good nights, the spectral observations of M31
GCs and GC candidates can be obtained. Based on these data, we can
study the properties of M31 GCs and GC candidates, such as
confirming GCs with radial velocities and determining metallicities
by measuring the strengths of various absorption features in the
integrated spectra (see Perrett et al. ~\cite{per02}, and references
there).

\begin{acknowledgements}
We are indebted to the referee for thoughtful comments and
insightful suggestions that improved this paper greatly. This work
is supported by the Chinese National Natural Science Foundation
through Grant Nos. 10873016, 10803007, 10473012, 10573020, 10633020,
10673012, and 10603006; and by National Basic Research Program of
China (973 Program) No. 2007CB815403.
\end{acknowledgements}

\clearpage

\begin{table}
\begin{center}
\caption[]{Comparison of the Coordinates from Caldwell et al.~(\cite{cald09}).  These five objects are from RBC V.3.5.}\label{Tab:coord2}
 \begin{tabular}{lcccc}
  \hline\hline\noalign{\smallskip}
  \multicolumn{1}{l}{} &\multicolumn{2}{c}{Caldwell et al.~(\cite{cald09})} &\multicolumn{2}{c}{RBC V.3.5} \\
\hline
  ID &  R.A.      &  DEC. & R.A. &    DEC.                \\
  &(J2000) &(J2000) &(J2000) &(J2000) \\
  \hline
  \multicolumn{1}{l}{(1)} &\multicolumn{1}{c}{(2)}&\multicolumn{1}{c}{(3)}&\multicolumn{1}{c}{(4)} &\multicolumn{1}{c}{(5)} \\
  \hline\noalign{\smallskip}
DAO23 & 00:38:54.19 & +40:26:33.9&  00:38:54.34 & +40:26:46.4 \\
V203  & 00:40:47.80 & +40:59:06.0&  00:40:48.62 & +40:59:11.4 \\
SH07  & 00:39:37.36 & +42:09:57.1 & 00:39:37.73 & +42:09:28.3 \\
V202  & 00:40:47.82 & +40:55:34.3 & 00:40:47.29 &+40:55:25.5 \\
DAO16 & 00:37:57.25 & +40:24:49.7 & 00:37:56.83 &+40:24:45.5 \\
  \noalign{\smallskip}\hline
 \end{tabular}
\end{center}
\end{table}

\begin{table}
  \begin{center}
    \caption[]{Comparison of the two BATC CCDs. \label{tab:ccd}}
    \begin{tabular}{lcc}
      \hline\hline\noalign{\smallskip}
      \multicolumn{1}{l}{Parameters} &\multicolumn{1}{c}{Old CCD} &\multicolumn{1}{c}{New E2V CCD}\\
      \multicolumn{1}{l}{(1)} &\multicolumn{1}{c}{(2)}&\multicolumn{1}{c}{(3)}\\
      \hline \noalign{\smallskip}
      Pixel Number   & $2048\times2048$ &4096$\times$4096 \\
      Spatial Scales &   $1.67\arcsec/\rm pixel$  & $1.33\arcsec/\rm pixel $ \\
      FOV            &  $58 \arcmin\times 58 \arcmin$&$92 \arcmin\times 92 \arcmin$\\
      Blue Quantum Efficiency ( at 4000 {\AA)} & $< 5\%$ & 92.2 $\%$ \\
      Red Quantum Efficiency ( at 8500 {\AA)} & $\sim 40\%$ & 59.4 $\%$ \\
      \noalign{\smallskip}\hline
    \end{tabular}
  \end{center}
\end{table}

\begin{table}
\begin{center}
\caption[]{Parameters of the BATC filters and statistics of
observations in M31-1 filed.\label{tab:T523ab}}
\begin{tabular}{lcccccc}
  \hline\hline \noalign{\smallskip}
  \multicolumn{1}{c}{Filter} &\multicolumn{1}{c}{C.W.$^a$ } &\multicolumn{1}{c}{Bandwidth$^b$ } &\multicolumn{1}{c}{Image Number$^c$} & \multicolumn{1}{c}{Exposure$^d$ } &\multicolumn{1}{c}{Calibration Error$^e$ }&\multicolumn{1}{c}{Limiting Mags~(S/N=5)$^f$}\\
  \multicolumn{1}{c}{} &\multicolumn{1}{c}{({\AA})} &\multicolumn{1}{c}{({\AA})} &\multicolumn{1}{c}{}&\multicolumn{1}{c}{(hr)}& \multicolumn{1}{c}{(MAG)}& \multicolumn{1}{c}{(MAG)}\\
  \hline
  \multicolumn{1}{l}{(1)} &\multicolumn{1}{c}{(2)}&\multicolumn{1}{c}{(3)}&\multicolumn{1}{c}{(4)} &\multicolumn{1}{c}{(5)} &\multicolumn{1}{c}{(6)}&\multicolumn{1}{c}{(7)}\\
  \hline\noalign{\smallskip}
$a$&   3360 & 222 &  6  & 1.5 &  0.018 &19.88\\
$b$&   3890 & 187 &  6  & 2.0 &  0.013 &18.43\\
  \noalign{\smallskip}\hline
\end{tabular}
\end{center}
\tablenotes{a}{0.7\textwidth}{Central wavelength for each BATC filer
in {\AA}.}\\
\tablenotes{b}{0.7\textwidth}{Bandwidth of the filter in {\AA}.}\\
\tablenotes{c}{0.7\textwidth}{The number of images combined to
increase the signal-to-noise ratios in each BATC filer.}\\
\tablenotes{d}{0.7\textwidth}{Total exposure time in
hours.}\\
\tablenotes{e}{0.7\textwidth}{The zero-point errors in magnitude
for each filter obtained from the standard stars.}\\
\tablenotes{f}{0.7\textwidth}{The limiting magnitudes for S/N=5.}\\
\end{table}

\begin{table}
\begin{center}
\caption[]{BATC Obervational Log for M31-2 -- M31-7 Fileds.
\label{tab:obs}}
\begin{tabular}{lcccccc}
\hline\hline \noalign{\smallskip}
\multicolumn{1}{l}{Field} &\multicolumn{1}{c}{Filter} &\multicolumn{1}{c}{Central Wavetlngth} &\multicolumn{1}{c}{Bandwidth} & \multicolumn{1}{c}{Image Number} &\multicolumn{1}{c}{Exposure} &\multicolumn{1}{c}{Limiting Mags~(S/N=5)}\\
\multicolumn{1}{l}{} &\multicolumn{1}{c}{} &\multicolumn{1}{c}{({\AA})} &\multicolumn{1}{c}{({\AA})}&\multicolumn{1}{c}{}&  \multicolumn{1}{c}{(hr)}& \multicolumn{1}{c}{(MAG)}\\
\hline
\multicolumn{1}{l}{(1)} &\multicolumn{1}{c}{(2)}&\multicolumn{1}{c}{(3)}&\multicolumn{1}{c}{(4)} &\multicolumn{1}{c}{(5)} &\multicolumn{1}{c}{(6)}&\multicolumn{1}{c}{(7)}\\
\hline \noalign{\smallskip}
 M31-2& $a$& 3360 & 222 & 7&  2.1 & 19.63\\
  ... & $b$& 3890 & 187 & 7&  2.1 & 19.96\\
  ... & $c$& 4210 & 185 & 4&  0.8 & 20.05\\
  ... & $d$& 4550 & 222 & 9&  3.0 & 20.09\\
  ... & $e$& 4920 & 225 & 3&  1.0 & 19.53\\
  ... & $f$& 5270 & 211 & 3&  1.0 & 18.60\\
  ... & $g$& 5795 & 176 & 3&  1.0 & 19.32\\
  ... & $h$& 6075 & 190 & 3&  1.0 & 19.29\\
  ... & $i$& 6660 & 312 & 5&  1.7 & 20.47\\
  ... & $j$& 7050 & 121 & 6&  2.0 & 19.25\\
  ... & $k$& 7490 & 125 & 4&  1.3 & 18.83\\
  ... & $m$& 8020 & 179 & 3&  1.0 & 19.12\\
  ... & $n$& 8480 & 152 & 5&  1.7 & 18.10\\
  ... & $o$& 9190 & 194 & 3&  1.0 & 18.53\\
  ... & $p$& 9745 & 188 & 5&  1.7 & 16.83\\
 M31-3& $a$& 3360 & 222 & 7&  2.1 & 17.55\\
  ... & $b$& 3890 & 187 & 6&  1.8 & 18.43\\
  ... & $c$& 4210 & 185 & 4&  1.0 & 20.08\\
  ... & $d$& 4550 & 222 & 3&  1.0 & 18.62\\
  ... & $e$& 4920 & 225 & 4&  1.3 & 19.76\\
  ... & $f$& 5270 & 211 & 3&  1.0 & 19.75\\
  ... & $g$& 5795 & 176 & 3&  1.0 & 19.50\\
  ... & $h$& 6075 & 190 & 3&  1.0 & 19.56\\
  ... & $i$& 6660 & 312 & 3&  1.0 & 18.10\\
  ... & $j$& 7050 & 121 & 3&  1.0 & 18.12\\
  ... & $k$& 7490 & 125 & 3&  1.0 & 19.47\\
  ... & $m$& 8020 & 179 & 3&  1.0 & 18.64\\
  ... & $n$& 8480 & 152 & 8&  2.7 & 17.89\\
  ... & $o$& 9190 & 194 & 4&  1.3 & 17.81\\
  ... & $p$& 9745 & 188 & 7&  2.3 & 17.92\\
 M31-4& $a$& 3360 & 222 & 6&  1.8 & 17.41\\
  ... & $b$& 3890 & 187 & 4&  1.1 & 17.47\\
  ... & $c$& 4210 & 185 & 6&  1.5 & 18.02\\
  ... & $d$& 4550 & 222 & 3&  1.0 & 20.03\\
  ... & $e$& 4920 & 225 & 3&  1.0 & 20.25\\
  ... & $f$& 5270 & 211 & 3&  1.0 & 19.49\\
  ... & $g$& 5795 & 176 & 3&  1.0 & 19.93\\
  ... & $h$& 6075 & 190 & 4&  1.3 & 20.30\\
  ... & $i$& 6660 & 312 & 3&  1.0 & 19.55\\
  ... & $j$& 7050 & 121 & 3&  1.0 & 19.57\\
  ... & $k$& 7490 & 125 & 3&  1.0 & 19.40\\
  ... & $m$& 8020 & 179 & 3&  1.0 & 18.93\\
  ... & $n$& 8480 & 152 & 5&  1.7 & 17.68\\
 \noalign{\smallskip}\hline
\end{tabular}
\end{center}
\end{table}
\addtocounter{table}{-1}
\begin{table}
\begin{center}
\caption[]{Continued- \label{tab:obs2}}
\begin{tabular}{lcccccc}
\hline\hline \noalign{\smallskip}
\multicolumn{1}{l}{Field} &\multicolumn{1}{c}{Filter} &\multicolumn{1}{c}{Central Wavetlngth} &\multicolumn{1}{c}{Bandwidth} & \multicolumn{1}{c}{Image Number} &\multicolumn{1}{c}{Exposure} &\multicolumn{1}{c}{Limiting Mags~(S/N=5)}\\
\multicolumn{1}{l}{} &\multicolumn{1}{c}{} &\multicolumn{1}{c}{({\AA})} &\multicolumn{1}{c}{({\AA})}&\multicolumn{1}{c}{}&  \multicolumn{1}{c}{(hr)}& \multicolumn{1}{c}{(MAG)}\\
\hline
\multicolumn{1}{l}{(1)} &\multicolumn{1}{c}{(2)}&\multicolumn{1}{c}{(3)}&\multicolumn{1}{c}{(4)} &\multicolumn{1}{c}{(5)} &\multicolumn{1}{c}{(6)}&\multicolumn{1}{c}{(7)}\\
\hline \noalign{\smallskip}
  ... & $o$& 9190 & 194 & 7&  2.3 & 18.95\\
  ... & $p$& 9745 & 188 & 3&  1.0 & 16.71\\
 M31-5& $a$& 3360 & 222 & 6&  2.0 & 16.54\\
  ... & $b$& 3890 & 187 & 8&  2.5 & 17.03\\
  ... & $c$& 4210 & 185 & 7&  1.8 & 20.50\\
  ... & $d$& 4550 & 222 & 3&  1.0 & 20.19\\
  ... & $e$& 4920 & 225 & 3&  1.0 & 19.49\\
  ... & $f$& 5270 & 211 & 3&  1.0 & 20.02\\
  ... & $g$& 5795 & 176 & 3&  1.0 & 20.03\\
  ... & $h$& 6075 & 190 & 3&  1.0 & 20.27\\
  ... & $i$& 6660 & 312 & 3&  1.0 & 20.07\\
  ... & $j$& 7050 & 121 & 3&  1.0 & 19.51\\
  ... & $k$& 7490 & 125 & 3&  1.0 & 19.35\\
  ... & $m$& 8020 & 179 & 3&  1.0 & 18.73\\
  ... & $n$& 8480 & 152 & 6&  2.0 & 18.21\\
  ... & $o$& 9190 & 194 & 3&  1.0 & 18.22\\
  ... & $p$& 9745 & 188 & 6&  2.0 & 18.33\\
 M31-6& $a$& 3360 & 222 & 7&  2.1 & 17.43\\
  ... & $b$& 3890 & 187 & 7&  2.1 & 17.58\\
  ... & $c$& 4210 & 185 & 7&  1.8 & 17.70\\
  ... & $d$& 4550 & 222 & 6&  2.0 & 16.07\\
  ... & $e$& 4920 & 225 & 4&  1.3 & 19.62\\
  ... & $f$& 5270 & 211 & 3&  1.0 & 19.69\\
  ... & $g$& 5795 & 176 & 3&  1.0 & 19.22\\
  ... & $h$& 6075 & 190 & 3&  1.0 & 19.25\\
  ... & $i$& 6660 & 312 & 6&  2.0 & 19.83\\
  ... & $j$& 7050 & 121 & 3&  1.0 & 18.97\\
  ... & $k$& 7490 & 125 & 3&  1.0 & 17.24\\
  ... & $m$& 8020 & 179 & 3&  1.0 & 17.95\\
  ... & $n$& 8480 & 152 & 5&  1.7 & 17.50\\
  ... & $o$& 9190 & 194 & 3&  1.0 & 17.59\\
  ... & $p$& 9745 & 188 & 6&  2.0 & 18.22\\
 M31-7& $a$& 3360 & 222 & 6&  2.0 & 19.66\\
  ... & $b$& 3890 & 187 & 6&  2.0 & 20.26\\
  ... & $c$& 4210 & 185 & 3&  0.8 & 15.49\\
  ... & $d$& 4550 & 222 & 3&  1.0 & 19.86\\
  ... & $e$& 4920 & 225 & 3&  1.0 & 20.16\\
  ... & $f$& 5270 & 211 & 3&  1.0 & 20.06\\
  ... & $g$& 5795 & 176 & 3&  1.0 & 19.24\\
  ... & $h$& 6075 & 190 & 3&  1.0 & 19.37\\
  ... & $i$& 6660 & 312 & 3&  1.0 & 19.99\\
  ... & $j$& 7050 & 121 & 5&  1.7 & 20.09\\
  ... & $k$& 7490 & 125 & 3&  1.0 & 17.73\\
 \noalign{\smallskip}\hline
\end{tabular}
\end{center}
\end{table}
\addtocounter{table}{-1}
\begin{table}
\begin{center}
\caption[]{Continued- \label{tab:obs3}}
\begin{tabular}{lcccccc}
\hline\hline \noalign{\smallskip}
\multicolumn{1}{l}{Field} &\multicolumn{1}{c}{Filter} &\multicolumn{1}{c}{Central Wavetlngth} &\multicolumn{1}{c}{Bandwidth} & \multicolumn{1}{c}{Image Number} &\multicolumn{1}{c}{Exposure} &\multicolumn{1}{c}{Limiting Mags~(S/N=5)}\\
\multicolumn{1}{l}{} &\multicolumn{1}{c}{} &\multicolumn{1}{c}{({\AA})} &\multicolumn{1}{c}{({\AA})}&\multicolumn{1}{c}{}&  \multicolumn{1}{c}{(hr)}& \multicolumn{1}{c}{(MAG)}\\
\hline
\multicolumn{1}{l}{(1)} &\multicolumn{1}{c}{(2)}&\multicolumn{1}{c}{(3)}&\multicolumn{1}{c}{(4)} &\multicolumn{1}{c}{(5)} &\multicolumn{1}{c}{(6)}&\multicolumn{1}{c}{(7)}\\
\hline \noalign{\smallskip}
  ... & $m$& 8020 & 179 & 3&  1.0 & 18.08\\
  ... & $n$& 8480 & 152 & 6&  2.0 & 18.28\\
  ... & $o$& 9190 & 194 & 6&  2.0 & 18.10\\
  ... & $p$& 9745 & 188 & 6&  2.0 & 17.92\\
 \noalign{\smallskip}\hline
\end{tabular}
\end{center}
\end{table}

\begin{table}
\begin{center}
\caption[]{Spectral energy distributions for 30 objects in M31
field. \label{tab:batc}}\tabcolsep=4pt
\begin{tabular}{lccccccccccccccc}
\hline\hline \noalign{\smallskip}
\multicolumn{1}{l}{ID} &\multicolumn{1}{c}{$a$} & \multicolumn{1}{c}{$b$} &\multicolumn{1}{c}{$c$} &\multicolumn{1}{c}{$d$}  &\multicolumn{1}{c}{$e$}  &\multicolumn{1}{c}{$f$}  &\multicolumn{1}{c}{$g$}  &\multicolumn{1}{c}{$h$}   &\multicolumn{1}{c}{$i$}  &\multicolumn{1}{c}{$j$}   &\multicolumn{1}{c}{$k$}   &\multicolumn{1}{c}{$m$}  &\multicolumn{1}{c}{$n$}  &\multicolumn{1}{c}{$o$}  &\multicolumn{1}{c}{$p$}\\
\multicolumn{1}{l}{(1)} &\multicolumn{1}{c}{(2)} &\multicolumn{1}{c}{(3)} &\multicolumn{1}{c}{(4)} &\multicolumn{1}{c}{(5)} &\multicolumn{1}{c}{(6)} &\multicolumn{1}{c}{(7)} &\multicolumn{1}{c}{(8)} &\multicolumn{1}{c}{(9)}&\multicolumn{1}{c}{(10)} &\multicolumn{1}{c}{(11)}  &\multicolumn{1}{c}{(12)} &\multicolumn{1}{c}{(13)}  &\multicolumn{1}{c}{(14)} &\multicolumn{1}{c}{(15)}&\multicolumn{1}{c}{(16)}\\
\hline \noalign{\smallskip}
B189D    & 18.98& 18.31& 18.18& 17.90& 18.00& 18.01& 17.91& 17.87& 17.78& 17.74& 17.96& 17.85& 18.15& 17.95& 17.78\\
         & 0.106& 0.050& 0.064& 0.088& 0.061& 0.058& 0.066& 0.062& 0.053& 0.068& 0.144& 0.098& 0.169& 0.262& 0.260\\
B193D    &  ... & 19.03& 18.64& 18.15& 18.05& 17.86& 17.51& 17.66& 17.40& 17.39& 17.33& 17.03& 17.36& 17.41& 17.43\\
         &  ... & 0.150& 0.098& 0.174& 0.192& 0.157& 0.164& 0.131& 0.105& 0.135& 0.150& 0.112& 0.184& 0.216& 0.268\\
G289     & 15.70& 14.55& 13.70& 13.12& 12.56& 12.48& 12.15& 12.07& 11.94& 11.75& 11.61& 11.63& 11.73& 11.68& 11.64\\
         & 0.007& 0.005& 0.002& 0.001& 0.002& 0.002& 0.002& 0.002& 0.001& 0.002& 0.002& 0.001& 0.002& 0.001& 0.002\\
G295     & 13.80& 12.91& 12.39& 11.90& 11.72& 11.70& 11.55& 11.56& 11.55& 11.36& 11.28& 11.34& 11.48& 11.53& 11.55\\
         & 0.003& 0.001& 0.001& 0.001& 0.001& 0.001& 0.001& 0.001& 0.001& 0.002& 0.002& 0.001& 0.002& 0.001& 0.002\\
NB34     &  ... & 20.02&  ... & 18.88&  ... & 17.76&  ... &  ... & 17.20& 17.47&  ... &  ... &  ... & 17.10&  ... \\
         &  ... & 0.454&  ... & 0.366&  ... & 0.160&  ... &  ... & 0.179& 0.300&  ... &  ... &  ... & 0.346&  ... \\
NB61     &  ... &  ... &  ... &  ... &  ... &  ... &  ... &  ... & 18.48& 17.70& 16.78& 16.69& 16.47& 16.12& 16.07\\
         &  ... &  ... &  ... &  ... &  ... &  ... &  ... &  ... & 0.760& 0.438& 0.225& 0.245& 0.232& 0.175& 0.177\\
AU008    &  ... & 19.64&  ... & 17.85& 17.52& 17.44& 17.99& 17.40& 17.02& 17.01&  ... &  ... &  ... &  ... &  ... \\
         &  ... & 0.313&  ... & 0.220& 0.232& 0.250& 0.568& 0.369& 0.306& 0.363&  ... &  ... &  ... &  ... &  ... \\
AU010    & 19.99& 19.10& 19.07& 18.09& 17.73& 17.53& 17.18& 16.95& 16.53& 16.32& 15.98& 15.77&  ... & 15.46& 15.46\\
         & 0.693& 0.637& 1.424& 0.867& 0.841& 0.758& 0.776& 0.657& 0.498& 0.457& 0.401& 0.372&  ... & 0.350& 0.362\\
DAO16    & 20.69& 19.55&  ... & 18.54&  ... &  ... & 18.38& 18.33&  ... &  ... &  ... &  ... &  ... &  ... &  ... \\
         & 0.342& 0.068&  ... & 0.122&  ... &  ... & 0.073& 0.056&  ... &  ... &  ... &  ... &  ... &  ... &  ... \\
DAO23    & 21.10& 20.49& 20.11& 19.41& 19.45& 19.47& 19.16& 19.18& 19.29& 19.13& 19.48& 19.32& 19.74& 19.11& 19.93\\
         & 0.345& 0.119& 0.161& 0.190& 0.116& 0.075& 0.096& 0.102& 0.079& 0.093& 0.320& 0.179& 0.414& 0.362& 1.132\\
DAO30    & 20.45& 19.34& 18.93& 18.37& 18.63& 18.39& 18.09& 17.99& 17.85& 17.83& 17.90& 17.72& 17.89& 17.95& 18.32\\
         & 0.374& 0.088& 0.094& 0.107& 0.083& 0.049& 0.057& 0.043& 0.031& 0.043& 0.116& 0.049& 0.106& 0.190& 0.359\\
DAO40    & 18.22& 17.55& 17.80& 17.87& 17.64&  ... & 17.94& 17.92& 16.67&  ... & 17.66& 17.61& 17.78& 17.19& 17.35\\
         & 0.058& 0.031& 0.051& 0.092& 0.050&  ... & 0.076& 0.070& 0.025&  ... & 0.125& 0.072& 0.127& 0.129& 0.193\\
DAO46    & 20.13& 19.54& 19.28& 18.61& 18.84& 18.70& 18.55& 18.39& 18.24& 18.25& 18.11& 18.13& 18.06& 18.16& 18.23\\
         & 0.260& 0.100& 0.133& 0.133& 0.097& 0.060& 0.070& 0.056& 0.039& 0.052& 0.140& 0.071& 0.125& 0.241& 0.376\\
DAO47    & 20.24& 19.23& 19.03& 18.60& 18.90& 18.84& 18.90& 19.01& 18.76& 18.64& 18.72& 18.46& 18.48& 18.09& 18.08\\
         & 0.319& 0.094& 0.119& 0.136& 0.115& 0.083& 0.133& 0.125& 0.104& 0.125& 0.258& 0.127& 0.195& 0.273& 0.325\\
DAO51    & 19.65& 18.51& 18.22& 17.44& 17.17& 16.94& 16.73& 16.52& 16.31& 16.24& 16.14& 15.96& 16.07& 15.92& 15.74\\
         & 0.164& 0.035& 0.043& 0.053& 0.028& 0.017& 0.020& 0.015& 0.009& 0.013& 0.032& 0.013& 0.025& 0.043& 0.052\\
DAO53    &  ... & 21.07& 20.12& 19.81& 19.40& 19.08& 18.70& 18.58& 18.24& 18.24& 18.11& 17.97& 17.85& 17.86& 17.79\\
         &  ... & 0.426& 0.303& 0.422& 0.170& 0.088& 0.099& 0.071& 0.039& 0.074& 0.145& 0.060& 0.098& 0.200& 0.255\\
DAO54    & 21.05& 20.30& 19.78& 19.12& 18.69& 18.61& 18.20& 18.13& 17.81& 17.81& 17.63& 17.63& 17.46& 17.36& 17.18\\
         & 0.593& 0.190& 0.204& 0.195& 0.082& 0.051& 0.058& 0.045& 0.023& 0.036& 0.087& 0.041& 0.059& 0.118& 0.146\\
DAO69    & 18.75& 17.91& 17.70& 17.83& 18.25& 17.77& 17.84& 18.15& 17.68& 17.92& 18.25& 17.55& 18.03& 17.99& 17.67\\
         & 0.119& 0.043& 0.039& 0.045& 0.093& 0.100& 0.063& 0.099& 0.046& 0.070& 0.143& 0.055& 0.128& 0.184& 0.246\\
DAO84    &  ... & 19.81& 19.42& 19.25& 19.26& 19.39& 19.11& 18.75& 18.48& 18.89& 18.71& 18.50& 18.54& 18.30& 18.69\\
         &  ... & 0.294& 0.070& 0.186& 0.145& 0.137& 0.140& 0.077& 0.074& 0.147& 0.168& 0.108& 0.266& 0.168& 1.008\\
V202     & 20.42& 19.04& 18.98& 19.03& 19.13&  ... & 19.38& 19.67&  ... &  ... &  ... &  ... &  ... &  ... &  ... \\
         & 0.286& 0.076& 0.087& 0.115& 0.169&  ... & 0.344& 0.474&  ... &  ... &  ... &  ... &  ... &  ... &  ... \\
V203     & 19.07& 18.98& 19.18& 19.38& 17.23&  ... &  ... &  ... & 17.07&  ... &  ... &  ... &  ... & 18.28&  ... \\
         & 0.075& 0.251& 0.083& 0.103& 0.016&  ... &  ... &  ... & 0.023&  ... &  ... &  ... &  ... & 0.126&  ... \\
V226     & 20.82& 19.34& 20.01& 19.84& 18.66& 19.54& 19.60& 19.84& 17.14& 19.25& 19.44& 19.60&  ... & 17.57&  ... \\
         & 0.339& 0.293& 0.086& 0.083& 0.034& 0.084& 0.105& 0.112& 0.019& 0.127& 0.174& 0.216&  ... & 0.064&  ... \\
 \noalign{\smallskip}\hline
\end{tabular}
\end{center}
\end{table}

\addtocounter{table}{-1}
\begin{table}
\begin{center}
\caption[]{Continued- \label{tab:batc2}}\tabcolsep=4pt
\begin{tabular}{lccccccccccccccc}
\hline\hline \noalign{\smallskip}
\multicolumn{1}{l}{ID} &\multicolumn{1}{c}{$a$} & \multicolumn{1}{c}{$b$} &\multicolumn{1}{c}{$c$} &\multicolumn{1}{c}{$d$}  &\multicolumn{1}{c}{$e$}  &\multicolumn{1}{c}{$f$}  &\multicolumn{1}{c}{$g$}  &\multicolumn{1}{c}{$h$}   &\multicolumn{1}{c}{$i$}  &\multicolumn{1}{c}{$j$}   &\multicolumn{1}{c}{$k$}   &\multicolumn{1}{c}{$m$}  &\multicolumn{1}{c}{$n$}  &\multicolumn{1}{c}{$o$}  &\multicolumn{1}{c}{$p$}\\
\multicolumn{1}{l}{(1)} &\multicolumn{1}{c}{(2)} &\multicolumn{1}{c}{(3)} &\multicolumn{1}{c}{(4)} &\multicolumn{1}{c}{(5)} &\multicolumn{1}{c}{(6)} &\multicolumn{1}{c}{(7)} &\multicolumn{1}{c}{(8)} &\multicolumn{1}{c}{(9)}&\multicolumn{1}{c}{(10)} &\multicolumn{1}{c}{(11)}  &\multicolumn{1}{c}{(12)} &\multicolumn{1}{c}{(13)}  &\multicolumn{1}{c}{(14)} &\multicolumn{1}{c}{(15)}&\multicolumn{1}{c}{(16)}\\
\hline \noalign{\smallskip}
V234     & 19.09& 18.55& 18.15& 17.96& 17.74& 17.59& 17.33& 17.33& 16.86& 17.08& 16.87& 16.67& 16.47& 16.44& 16.45\\
         & 0.077& 0.154& 0.038& 0.041& 0.053& 0.052& 0.063& 0.058& 0.057& 0.071& 0.074& 0.080& 0.080& 0.086& 0.095\\
V245     & 18.91& 19.24& 18.77& 19.26& 18.51& 19.90& 21.47&  ... & 17.13&  ... &  ... &  ... &  ... &  ... &  ... \\
         & 0.081& 0.291& 0.040& 0.051& 0.012& 0.025& 0.059&  ... & 0.009&  ... &  ... &  ... &  ... &  ... &  ... \\
BA28     &  ... & 20.67& 20.17& 19.68& 19.68& 19.48& 18.65& 18.58& 18.32& 18.35& 17.80& 17.94& 17.80& 17.74& 18.69\\
         &  ... & 0.453& 0.285& 0.222& 0.269& 0.649& 0.129& 0.090& 0.045& 0.096& 0.086& 0.071& 0.109& 0.145& 0.665\\
SH07     & 18.49& 17.57& 17.66& 17.41& 17.05& 16.85& 16.63& 16.55& 16.34& 16.35& 16.23& 16.02& 16.03& 15.98& 15.88\\
         & 0.075& 0.024& 0.013& 0.056& 0.030& 0.017& 0.026& 0.013& 0.020& 0.037& 0.016& 0.015& 0.029& 0.045& 0.045\\
BH10     &  ... & 20.49& 20.31& 19.29& 19.29& 19.39& 18.91& 18.70& 18.69& 18.77& 18.85& 18.60& 19.06& 18.43& 18.70\\
         &  ... & 0.300& 0.422& 0.263& 0.193& 0.182& 0.173& 0.130& 0.119& 0.156& 0.362& 0.177& 0.401& 0.393& 0.644\\
B515     &  ... & 19.93& 19.77& 19.16& 18.70& 18.65& 18.37& 18.20&  ... & 17.87& 17.87& 17.73&  ... & 17.66&  ... \\
         &  ... & 0.394& 0.074& 0.042& 0.037& 0.033& 0.052& 0.036&  ... & 0.057& 0.072& 0.076&  ... & 0.074&  ... \\
B521     &  ... & 19.72& 19.71& 19.87& 19.23& 18.96& 19.00& 19.00& 18.68& 18.67& 18.61& 18.43& 18.30& 17.90& 17.72\\
         &  ... & 0.130& 0.214& 0.379& 0.158& 0.101& 0.141& 0.130& 0.094& 0.119& 0.228& 0.109& 0.154& 0.197& 0.235\\
B524     &  ... &  ... & 19.99& 20.03& 19.71& 19.88& 19.21& 19.28& 19.42& 18.91& 19.19& 18.95& 18.25& 18.49& 18.40\\
         &  ... &  ... & 0.274& 0.218& 0.160& 0.206& 0.136& 0.117& 0.205& 0.138& 0.235& 0.210& 0.148& 0.184& 0.248\\
 \noalign{\smallskip}\hline
\end{tabular}
\end{center}
\end{table}

\begin{table}
\begin{center}
\caption[]{The broadband $UBVRI$ magnitudes of our sample GCs and candidates
derived from the BATC 15-band photometry. \label{tab:broad}}
\begin{tabular}{lcccccccccc}
\hline\hline \noalign{\smallskip}
\multicolumn{1}{l}{ID} &\multicolumn{1}{c}{$U$} &\multicolumn{1}{c}{$\sigma_{U}$} & \multicolumn{1}{c}{$B$} & \multicolumn{1}{c}{$\sigma_{B}$} &\multicolumn{1}{c}{$V$} & \multicolumn{1}{c}{$\sigma_{V}$}&\multicolumn{1}{c}{$R$} & \multicolumn{1}{c}{$\sigma_{R}$} &\multicolumn{1}{c}{$I$}& \multicolumn{1}{c}{$\sigma_{I}$}\\
\multicolumn{1}{l}{(1)} &\multicolumn{1}{c}{(2)} &\multicolumn{1}{c}{(3)} &\multicolumn{1}{c}{(4)} &\multicolumn{1}{c}{(5)} &\multicolumn{1}{c}{(6)} &\multicolumn{1}{c}{(7)} &\multicolumn{1}{c}{(8)} &\multicolumn{1}{c}{(9)} &\multicolumn{1}{c}{(10)} &\multicolumn{1}{c}{(11)}\\
\hline \noalign{\smallskip}

B189D    & 17.87& 0.094& 18.07& 0.090& 18.01& 0.072& 17.88& 0.053& 17.92& 0.344\\
B193D    &  ... &  ... & 18.40& 0.180& 17.62& 0.177& 17.50& 0.105& 17.06& 0.318\\
G289     & 14.43& 0.008& 13.50& 0.001& 12.33& 0.002& 12.04& 0.001& 11.44& 0.002\\
G295     & 12.61& 0.002& 12.18& 0.001& 11.64& 0.001& 11.65& 0.001& 11.18& 0.002\\
NB34     &  ... &  ... &  ... &  ... &  ... &  ... & 17.31& 0.179&  ... &  ... \\
NB61     &  ... &  ... &  ... &  ... &  ... &  ... & 18.58& 0.760& 16.11& 0.273\\
AU008    &  ... &  ... &  ... &  ... & 18.05& 0.587& 17.13& 0.306&  ... &  ... \\
AU010    & 18.81& 0.903& 18.52& 0.940& 17.42& 0.843& 16.63& 0.498&  ... &  ... \\
DAO16    & 19.43& 0.247&  ... &  ... &  ... &  ... &  ... &  ... &  ... &  ... \\
DAO23    & 20.01& 0.275& 19.68& 0.195& 19.30& 0.105& 19.40& 0.079& 18.68& 0.939\\
DAO30    & 19.20& 0.276& 18.56& 0.111& 18.27& 0.061& 17.95& 0.031& 17.34& 0.329\\
DAO40    & 17.11& 0.054& 18.03& 0.093&  ... &  ... & 16.78& 0.025& 17.20& 0.210\\
DAO46    & 19.04& 0.214& 18.83& 0.138& 18.70& 0.075& 18.35& 0.039& 17.74& 0.373\\
DAO47    & 19.02& 0.245& 18.76& 0.141& 18.89& 0.142& 18.86& 0.104& 18.09& 0.386\\
DAO51    & 18.39& 0.119& 17.80& 0.054& 16.91& 0.021& 16.41& 0.009& 15.86& 0.060\\
DAO53    &  ... &  ... & 20.10& 0.429& 18.91& 0.106& 18.34& 0.039& 17.60& 0.280\\
DAO54    & 19.91& 0.464& 19.49& 0.201& 18.41& 0.062& 17.91& 0.023& 17.25& 0.164\\
DAO69    & 17.58& 0.096& 17.84& 0.050& 17.76& 0.078& 17.78& 0.046& 17.95& 0.271\\
DAO84    &  ... &  ... & 19.41& 0.189& 19.37& 0.149& 18.58& 0.074& 17.89& 0.768\\
V202     & 19.08& 0.215& 19.13& 0.122&  ... &  ... &  ... &  ... &  ... &  ... \\
V203     & 18.14& 0.308& 19.94& 0.105&  ... &  ... & 17.17& 0.023&  ... &  ... \\
V226     & 19.45& 0.423& 20.26& 0.085& 19.55& 0.115& 17.25& 0.019&  ... &  ... \\
V234     & 18.02& 0.193& 18.18& 0.043& 17.47& 0.068& 16.96& 0.057& 16.16& 0.124\\
V245     & 18.12& 0.356& 19.44& 0.052&  ... &  ... & 17.23& 0.009&  ... &  ... \\
BA28     &  ... &  ... & 19.91& 0.238& 18.99& 0.251& 18.42& 0.045& 16.80& 0.506\\
SH07     & 17.30& 0.059& 17.67& 0.056& 16.78& 0.027& 16.44& 0.020& 15.79& 0.059\\
BH10     &  ... &  ... & 19.64& 0.282& 19.19& 0.188& 18.80& 0.119& 18.39& 0.672\\
B515     &  ... &  ... & 19.52& 0.046& 18.57& 0.054&  ... &  ... &  ... &  ... \\
B521     &  ... &  ... & 20.11& 0.383& 19.04& 0.151& 18.78& 0.094& 18.01& 0.282\\
B524     &  ... &  ... & 20.22& 0.229& 19.45& 0.157& 19.53& 0.205& 18.08& 0.277\\
 \noalign{\smallskip}\hline
\end{tabular}
\end{center}
\end{table}

\label{lastpage}
\end{document}